# First-principles study of formic acid decomposition on single Pt atoms supported on heteroatom-doped graphene

Kazuma Sato, Norihito Sakaguchi, Yuji Kunisada*

*Center for Advanced Research of Energy and Materials, Faculty of Engineering, Hokkaido University, Kita 13 Nishi 8 Kita-ku, Sapporo, Hokkaido, 060-8628, Japan*

*E-mail: kunisada@eng.hokudai.ac.jp

Formic acid is a promising liquid hydrogen carrier, but its catalytic decomposition requires both high activity and selectivity toward dehydrogenation. Here, we investigated the catalytic activity and selectivity of formic acid decomposition on single Pt atoms supported on pristine and heteroatom-doped graphene using first-principles calculations based on the density functional theory. Reaction energy profiles reveal that single Pt atoms on P- and O-doped graphene show catalytic activity comparable to Pt(111), while all systems maintain strong dehydrogenation selectivity. These findings highlight doped graphene as a promising support for reducing precious metal usage in dehydrogenation catalysts.

The use of fossil fuel-derived energy, which emits large amounts of greenhouse gas, such as carbon dioxide, is a major cause of global warming. As a potential solution, renewable energy sources such as solar and wind power, which do not emit carbon dioxide during electricity generation, have attracted considerable attention. However, the output of solar and wind power generation fluctuates significantly depending on seasonal and weather conditions.[1,2] To address this issue, storing excess electricity in the form of chemical energy, such as hydrogen, has emerged as a vital complement to conventional batteries.[3-5]

Currently, the most widely adopted hydrogen storage technology is compressed gas cylinders, but these systems necessitate high-pressure handling and raise concerns regarding vessel safety. Another strategy involves storing hydrogen through reactions with metallic materials or intermetallic compounds to form metal hydrides; however, their gravimetric hydrogen density is typically limited (<3 mass%).[6] Magnesium, a lightweight metal, can form magnesium hydride with relatively high gravimetric hydrogen storage capacity (7.6 mass%), but the hydrogen desorption release process requires heating above 350°C.[7] To overcome these limitations, hydrogen storage using organic compounds, such as ammonia, methanol, methylcyclohexane, and particularly formic acid, has been explored.[3,8-10] Among these, liquid hydrogen storage using formic acid has attracted particular interest as a liquid hydrogen storage medium.[11] Although the hydrogen storage capacity of formic acid (4.4 mass%) falls short of the U.S. Department of Energy, its liquid state at room temperature ensures easy handling, and its gravimetric hydrogen density surpasses that of many hydrogen storage alloys. Therefore, formic acid is regarded as a promising candidate for next-generation liquid hydrogen carriers.

Precious metals such as Pt, Pd, Au, Ag, and Rh have been reported to exhibit high activity in the dehydrogenation reaction of formic acid.[12-16] However, these metals suffer from drawbacks including high production costs and limited proven reserves, creating a demand for reducing the amount of catalyst metal required. Furthermore, because the dehydrogenation of formic acid is accompanied by a competing dehydration pathway that produces $H_2O$ rather than $H_2$, thereby reducing hydrogen yield, improving reaction selectivity toward dehydrogenation remains a significant challenge. One promising approach to overcome these issues is to reduce catalyst size to the single-atom level, which achieves a higher specific surface area, an emerging quantum effect, and a larger support effect. However, isolated atoms generally tend to aggregate. To address, supporting single-atom catalysts on graphene doped with light elements as anchoring sites is widely studied to suppress aggregation and enhance catalyst activity and reaction selectivity through the suitable support effects.[17-20]

In this paper, we systematically investigated the decomposition of formic acid on Pt single atoms supported on light-element-doped graphene using first-principles calculations based on density functional theory (DFT). In particular, we examined the dehydrogenation and dehydration pathways of formic acid, with an emphasis on catalytic activity and reaction selectivity.

The calculations in this study were performed using the Vienna Ab Initio Simulation Package (VASP),[21–24] a first-principles calculation code employing plane-wave basis sets. For the exchange-correlation functional, we adopted the rev-vdW-DF, proposed by Hamada, which can accurately describe van der Waals interactions.[25] The cutoff energy for the plane-wave basis was set to 600 eV, and the Brillouin zone sampling employed a 4×4×1 Monkhorst-Pack grid.[26] These parameters ensured energy convergence within 0.01 eV per atom. Smearing was applied using a Gaussian function with $\sigma = 0.1$ eV. For periodic calculations, a 4×4 graphene supercell separated by a 20 Å vacuum layer, as shown in Fig. 1, was employed. For calculating isolated formic acid in vacuum, a cubic supercell of 20×20×20 Å$^3$ with 1×1×1 k-point sampling was employed. Structural relaxation was performed until residual forces converged to within 0.02 eV/Å. The activation barrier of each reaction step was evaluated using the Climbing Image Nudged Elastic Band (CI-NEB) method[27,28] with five intermediate images. To compensate for the dipole–dipole interaction between graphene layers, a dipole moment correction was incorporated.[29] Finally, the atomic structures were visualized using Visualization for Electronic and Structure Analysis (VESTA).[30]

We have previously reported the most stable adsorption configurations and adsorption energies of single Pt atoms on various light-element-doped graphene.[19,20] We considered B, N, O, Si, P, and S as dopants. All dopants introduce stronger bonding between the Pt-graphene substrate, which acts as the anchoring site. For most dopants, the dopant-top site was found to be the most stable, whereas the dopant-C bridge site was preferred in the case of B-doped graphene. Adsorption strength followed the order N < B < S < Si < P < O. In particular, P-, Si-, and S-doped graphene, which exhibit out-of-plane protrusions and dangling bonds through sp$^3$ hybrid orbital formation at the dopant-top site, stabilized Pt more effectively than planar dopants such as B and N, and O. We note that O-doped graphene shows the strongest adsorption among the considered dopants, although O-doped graphene has the planar structure. This is because the O dopants at three-fold sites form chemical bonds with only two C atoms, which introduces the dangling bond at the other C atoms. Based on these findings, four graphene supports were selected for further detailed analysis of formic acid decomposition: pristine graphene, N-doped graphene, which is the most easily doped, P-doped graphene, which exhibits the strongest adsorption among protruded structures, and O-doped graphene, which exhibits the strongest adsorption among planar doped systems.

In this study, we investigated two competing reactions involved in the decomposition of formic acid: dehydrogenation (HCOOH⇄H$_2$+CO$_2$) and dehydration (HCOOH⇄H$_2$O+CO). For the dehydrogenation, we considered two reaction pathways: the formate pathway, in which formate forms as a reaction intermediate, and the carboxyl pathway, in which a carboxyl group forms. For dehydration, we also considered two reaction pathways: the carboxyl pathway and the formyl pathway, in which a formyl group forms. We considered the following three adsorption states: (i) the initial adsorption state (IS), where formic acid is adsorbed on Pt supported on light-element-doped graphene; (ii) intermediate adsorption states (IM), where hydrogen atoms and formate, carboxyl, or formyl intermediates, decomposed from formic acid, on Pt supported on light-element-doped graphene; and (iii) the final adsorption states (FS), where either two hydrogen atoms and $CO_2$, or $H_2O$ and CO, are adsorbed on the Pt/graphene system. For IM, the initial structures included two configurations in the formate pathway: monodentate adsorption ($IM_{fm}$) and bidentate adsorption ($IM_{fb}$). In the carboxyl pathway, two configurations were considered: adsorption through the O atom ($IM_{cO}$) and adsorption through the C atom ($IM_{cC}$) of the carboxyl group. In the formyl pathway, one configuration ($IM_f$) was considered. The Atomic structures of these configurations are shown in Supplementary Fig. S1. For FS, as the initial structure, one configuration in dehydrogenation and two in dehydration were considered, owing to the different positions at which formic acid fragments in dehydration. All structures were relaxed to identify the most stable configurations. The origin of the relative energy $E_r$ for evaluating the energy profile was set to the sum of isolated Pt supported on light-element-doped graphene and formic acid, according to the following equation:

$$E_r = E_{complex} - (E_{Pt/gra} + E_{HCOOH}), \qquad (1)$$

where $E_{complex}$ and $E_{gra+Pt}$ are the total energy of single Pt atoms supported on light-element-doped graphene with and without adsorbates; $E_{HCOOH}$ is the total energy of isolated formic acid in vacuum.

Supplementary Table S1 summarizes the relative energies $E_r$ for each reaction stage in the decomposition of formic acid on single Pt atoms supported on pristine, N-doped, P-doped, and O-doped graphene. The relative energies of IS, IM, and FS decrease in the order of pristine > N-doped > P-doped > O-doped graphene. Based on these relative energies, the following pathways were examined. For pristine graphene: $IM_{fm}$, $IM_{fb}$, and $IM_{cC}$ in the dehydrogenation reaction; $IM_{cC}$ and $IM_f$ in the dehydration reaction. For N-doped graphene: $IM_{fb}$ and $IM_{cC}$ in dehydrogenation; $IM_{cC}$ and $IM_f$ in dehydration. For P-doped graphene: $IM_{fm}$, $IM_{fb}$, and $IM_{cC}$ in dehydrogenation; $IM_{cC}$ and $IM_f$ in dehydration. For O-doped graphene: $IM_{fm}$, $IM_{fb}$, and $IM_{cC}$ in dehydrogenation; $IM_{cC}$ and $IM_f$ in dehydration. In all cases, $IM_{cO}$ was excluded, as the structures were energetically unfavorable. We note that the energy difference between the two FS structures in the dehydration reaction was less than 0.2 eV; therefore, this difference does not affect the subsequent discussion.

Figure 2 shows the energy profiles of dehydrogenation and dehydration on single Pt atoms supported on pristine graphene, which presents the relative energy of each reaction step end transition states. In this study, the reaction process from IS to IM is referred to as step 1, while that from IM to FS is referred to as step 2. From Fig. 2(a), the activation barriers for steps 1 and 2 of the dehydrogenation pathway via $IM_{fm}$ were 0.68 and 1.16 eV, respectively; via $IM_{fb}$, 0.32 and 1.78 eV, respectively; and via $IM_{cC}$, 0.97 and 0.94 eV, respectively. The desorption of H₂ and CO₂ from Pt required 2.06 eV. Thus, the rate-limiting step in all dehydrogenation pathways on single Pt atoms on pristine graphene is the desorption of products. From Fig. 2(b), the activation barriers for steps 1 and 2 of the dehydration pathway via IM$_{cC}$ were 0.97 and 1.25 eV, respectively; and via IM$_f$, 0.78 and 0.77 eV, respectively. The desorption of H₂O and CO required 3.48 eV, indicating that the rate-limiting step in dehydration is also the desorption of products. Comparing the maximum activation barriers of the rate-limiting steps for dehydrogenation and dehydration, that of the dehydrogenation pathway is 1.42 eV lower than that of dehydration, suggesting superior reaction selectivity for dehydrogenation on single Pt atoms on pristine graphene.

Figure 3 shows the energy profiles of dehydrogenation and dehydration on single Pt atoms supported on N-doped graphene. As shown in Fig. 3(a), the activation barriers for steps 1 and 2 of the dehydrogenation pathway via IM$_{fb}$ were 0.34 and 1.16 eV, respectively; and via IM$_{cC}$, 0.83 and 1.03 eV, respectively. The desorption of H₂ and CO₂ from single Pt atoms required 1.66 eV, indicating that the rate-limiting step for all dehydrogenation pathways is the desorption of products. As shown in Fig. 3(b), the activation barriers for steps 1 and 2 of the dehydration pathway via IM$_{cC}$ were 0.83 and 1.66 eV, respectively; and via IM$_f$, 0.82 and 0.79 eV, respectively. The desorption of H₂O and CO required 3.14 eV, indicating that the rate-limiting step in all dehydration pathways is the desorption of products. Comparing the maximum activation barriers of the rate-limiting steps for dehydrogenation and dehydration, that of the dehydrogenation pathway is 1.48 eV lower than that of the dehydration pathway, suggesting superior reaction selectivity toward dehydrogenation on single Pt atoms on N-doped graphene.

Figure 4 shows the energy profiles of dehydrogenation and dehydration on single Pt atoms supported on P-doped graphene. From Fig. 4(a), the activation barriers for steps 1 and 2 of the dehydrogenation pathway via IM$_{fm}$ were 0.43 and 0.58 eV, respectively; via IM$_{fb}$, 0.42 and 1.07 eV, respectively; and via IM$_{cC}$, 0.81 and 1.64 eV, respectively. The desorption of H₂ and CO₂ from single Pt atoms required 0.92 eV. From these results, the most favorable pathway in the dehydrogenation reaction is the one proceeding via IM$_{fm}$, and the rate-limiting step corresponds to the product desorption step. From Fig. 4(b), the activation barriers for steps 1 and 2 of the dehydration pathway via IM$_{cC}$ were 0.81 and 0.02 eV, respectively; and via IM$_f$, 0.80 and 1.42 eV, respectively. The desorption of H₂O and CO required 2.41 eV, indicating that the rate-limiting step in all dehydration pathways is the desorption of products. Comparing the maximum activation barriers of the rate-limiting steps for dehydrogenation and dehydration, that of the dehydrogenation pathway is 1.49 eV lower than that of the dehydration pathway, suggesting superior reaction selectivity toward dehydrogenation on single Pt atoms on P-doped graphene.

Figure 5 shows the energy profiles of dehydrogenation and dehydration on single Pt atoms supported on O-doped graphene. From Fig. 5(a), the activation barriers for steps 1 and 2 of the dehydrogenation pathway via IM$_{fm}$ were 0.31 and 0.82 eV, respectively; via IM$_{fb}$, 0.45 and 1.10 eV, respectively; via $IM_{cC}$, 0.83 and 0.27 eV, respectively. The desorption of $H_2$ and $CO_2$ from single Pt atoms required 0.65 eV. From these results, the most favorable pathway in the dehydrogenation reaction is the one proceeding via $IM_{fm}$, and the rate-limiting step corresponds to step 2. As shown in Fig. 5(b), the activation barriers for steps 1 and 2 of the dehydration pathway via $IM_{cC}$ were 0.83 and 1.01 eV, respectively; via $IM_f$ were 1.19 and 0.86 eV, respectively. The desorption of $H_2O$ and CO required 2.20 eV, indicating that the rate-limiting step in all dehydration pathways is the desorption of products. Comparing the maximum activation barriers of the rate-limiting steps for dehydrogenation and dehydration, that of the dehydrogenation pathway is 1.38 eV lower than that of the dehydration pathway, suggesting superior reaction selectivity toward dehydrogenation on single Pt atoms on O-doped graphene.

When comparing the investigated supports, the maximum activation barrier for dehydrogenation was found to decrease in the order of pristine, N-doped, P-doped, and O-doped graphene, indicating that O-doping provides the most favorable reaction pathway and the highest catalytic activity. Notably, P- and O-doped graphene supports achieve both the increased specific surface area provided by single-atom dispersion and catalytic activity per active site comparable to that of the Pt(111) surface, whose maximum activation barrier of 0.81 eV.[16] Furthermore, since the maximum activation barrier for dehydration is consistently ~1.45 eV higher than that for dehydrogenation across all supports, all investigated systems exhibit favorable selectivity toward dehydrogenation. In addition, the observed trend in catalytic activity among the investigated systems is opposite to the order of adsorption energies of single Pt atoms on the supports, i.e., the bonding strength between Pt and the supports.[20] This is because, in almost all systems, highly active single Pt atoms make the product desorption step the rate-determining process. Indeed, the trend in product stability observed in this study is consistent with the previously reported stability of $H_2$ adsorption on single Pt atoms supported on these graphene. As discussed in ref. 30, these results indicate that heteroatom doping, which strengthens the binding between single Pt atoms and the supports, while simultaneously weakening the interaction between the adsorbate and the single Pt atoms, lowers the activation barrier for the desorption of $H_2$ and $CO_2$ and thereby enhances catalytic activity. In particular, along the dehydrogenation pathway via $IM_{fm}$ on O-deped graphene, the desorption energy of products became lower than the activation barrier of step 2, resulting in a change in the rate-limiting process. These findings suggest that further enhancement of catalytic performance may be achieved by considering more inert single-atom catalysts or supports with stronger anchoring sites.

In conclusion, doping graphene with N, P, or O was found to enhance the catalytic activity of single Pt atoms compared with pristine graphene supports. Among these dopants, O-doped graphene provided the highest catalytic activity. The obtained results demonstrate that introducing P or O into graphene supports can elevate the performance of single-atom catalysts to a level comparable to bulk Pt, thereby offering a promising strategy for reducing precious metal usage in formic acid dehydrogenation catalysts.

**Acknowledgments**

This work was supported by JSPS KAKENHI Grant Number JP23K04577 and the University Research Support System of Hokkaido Gas. This work has been done using the facilities of the Supercomputer Center, the Institute for Solid State Physics, the University of Tokyo.

# Figures

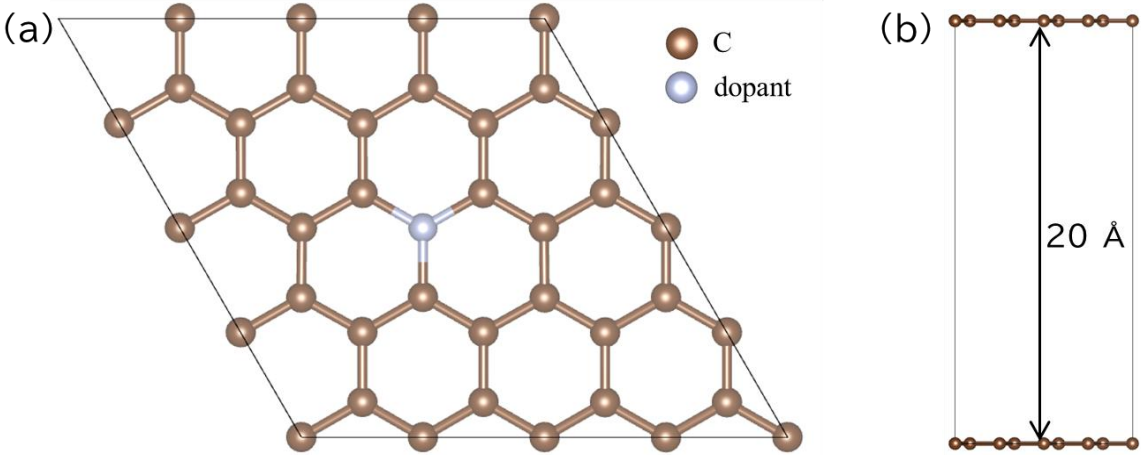

**Fig. 1.** (a) Top view and (b) side view of atomic structure of light-element-doped graphene. The black lines represent the supercell employed in this study.

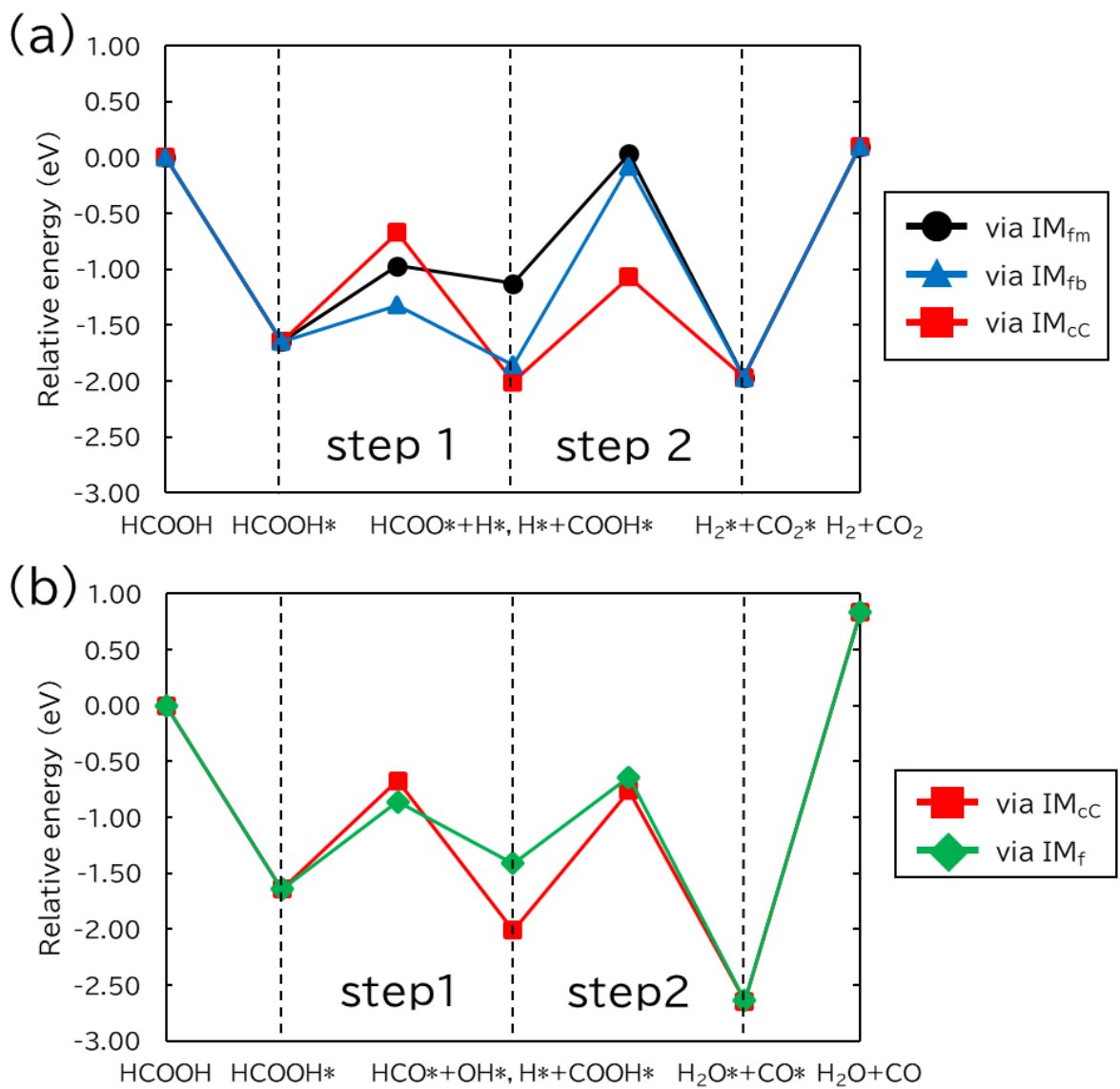

**Fig. 2.** Energy profiles for (a) dehydrogenation and (b) dehydration on single Pt atoms supported on pristine graphene. The asterisk denotes adsorbed states.

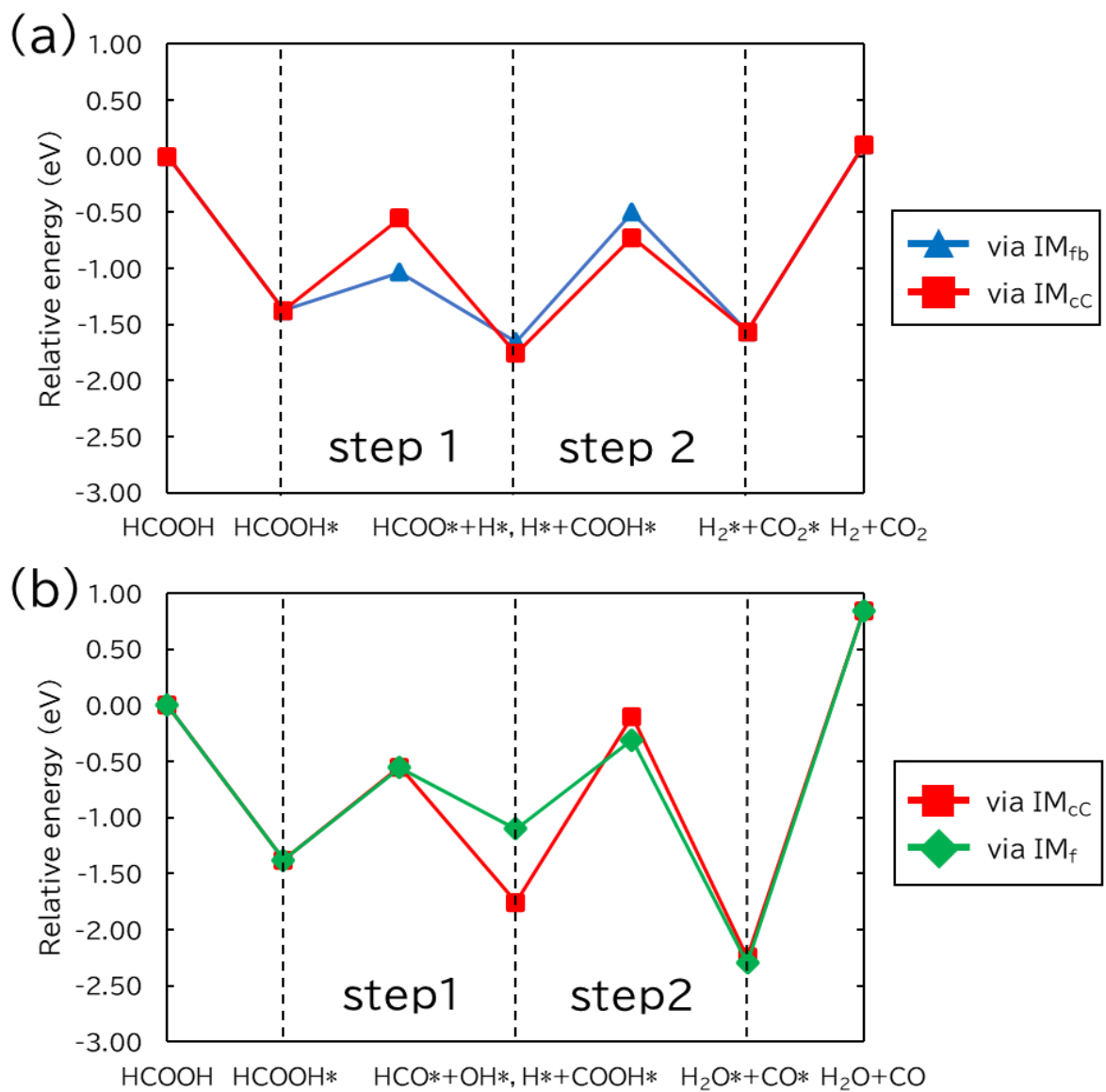

**Fig. 3.** Energy profiles for (a) dehydrogenation and (b) dehydration on single Pt atoms supported on N-doped graphene. The asterisk denotes adsorbed states.

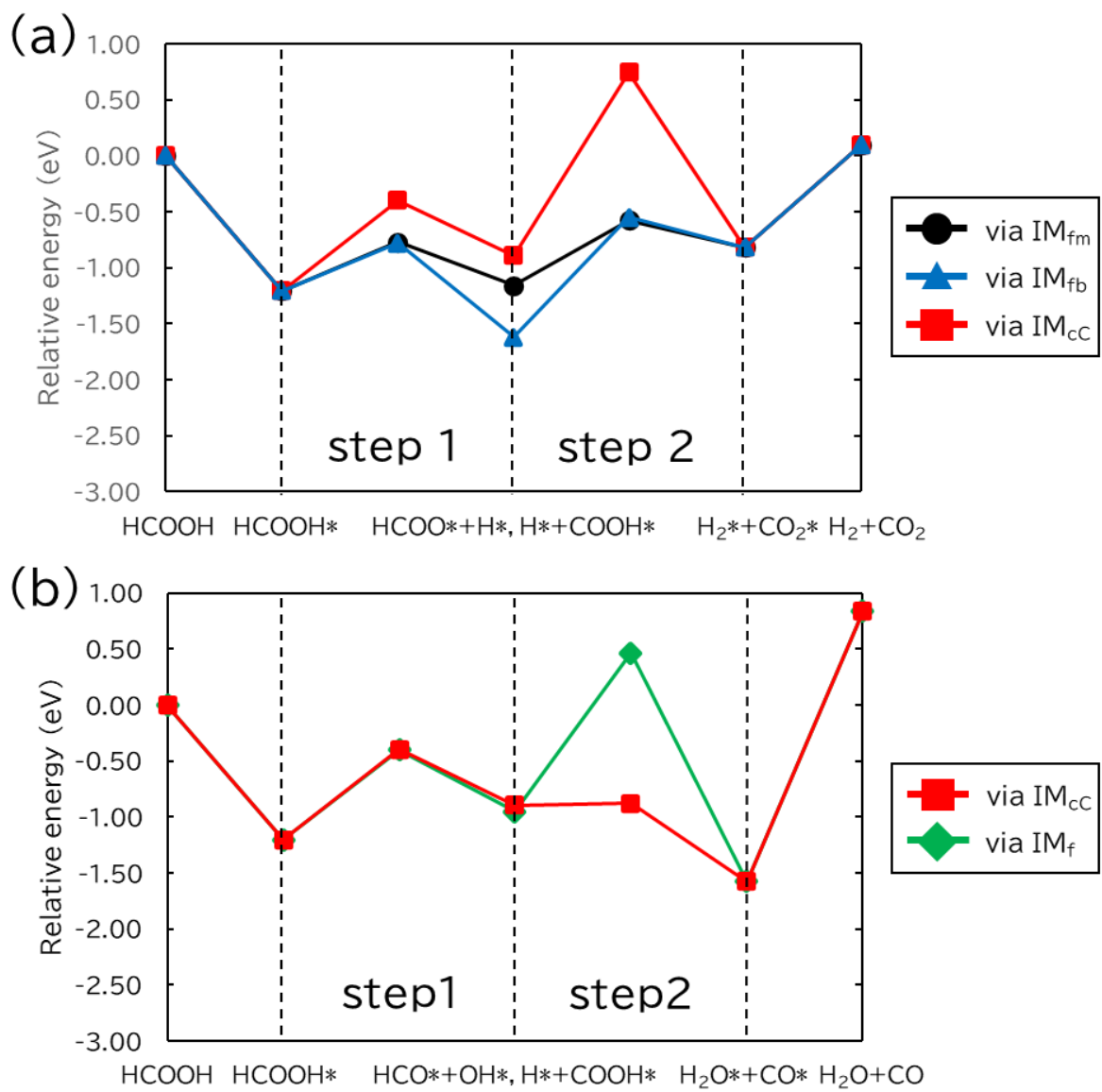

**Fig. 4.** Energy profiles for (a) dehydrogenation and (b) dehydration on single Pt atoms supported on P-doped graphene. The asterisk denotes adsorbed states.

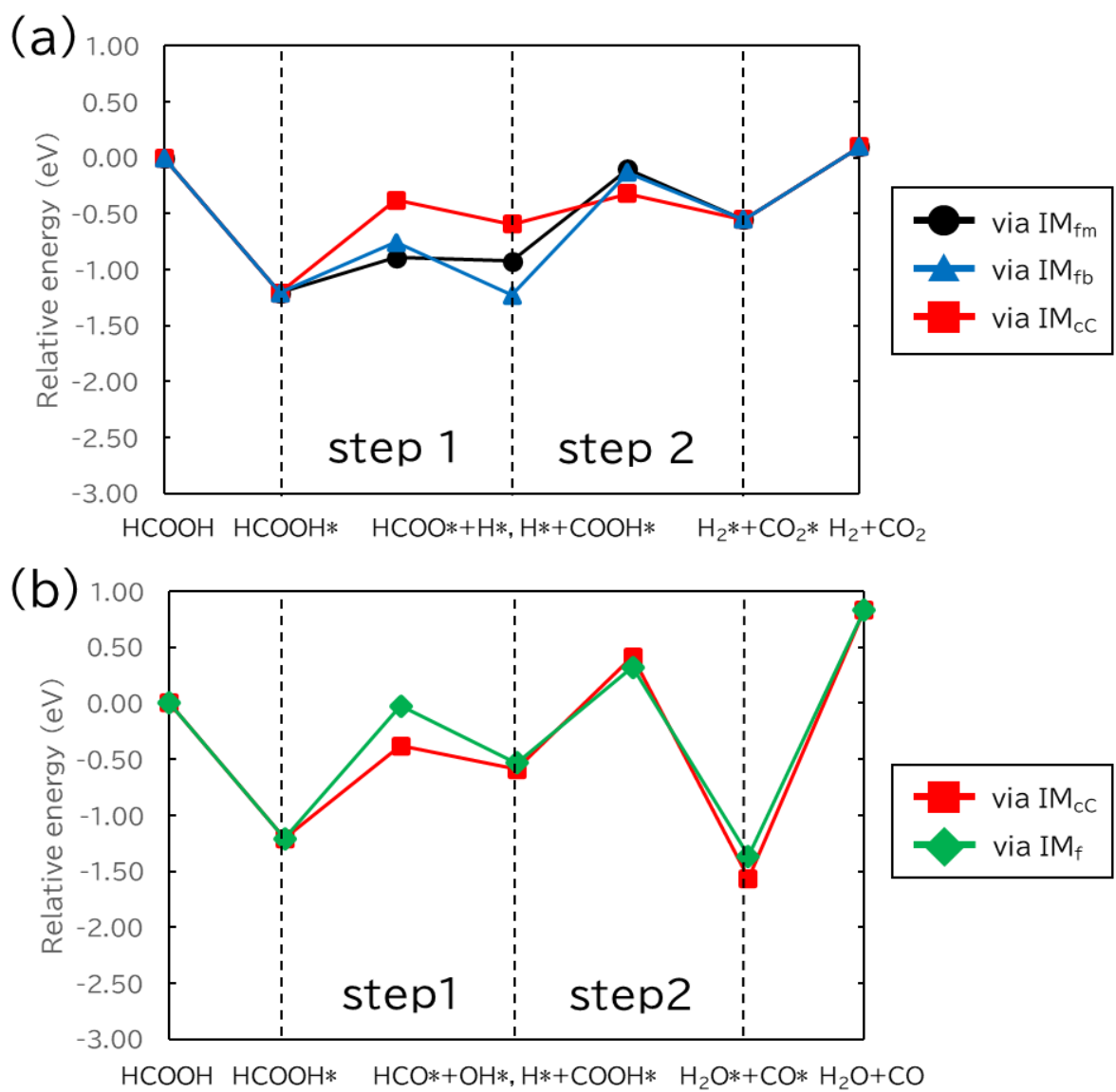

**Fig. 5.** Energy profiles for (a) dehydrogenation and (b) dehydration on single Pt atoms supported on O-doped graphene. The asterisk denotes adsorbed states.

# Supplementary data

# First-principles study of formic acid decomposition on single Pt atoms supported on heteroatom-doped graphene

Kazuma Sato, Norihito Sakaguchi, Yuji Kunisada*

*Center for Advanced Research of Energy and Materials, Faculty of Engineering, Hokkaido University, Kita 13 Nishi 8 Kita-ku, Sapporo, Hokkaido, 060-8628, Japan*

*E-mail: kunisada@eng.hokudai.ac.jp

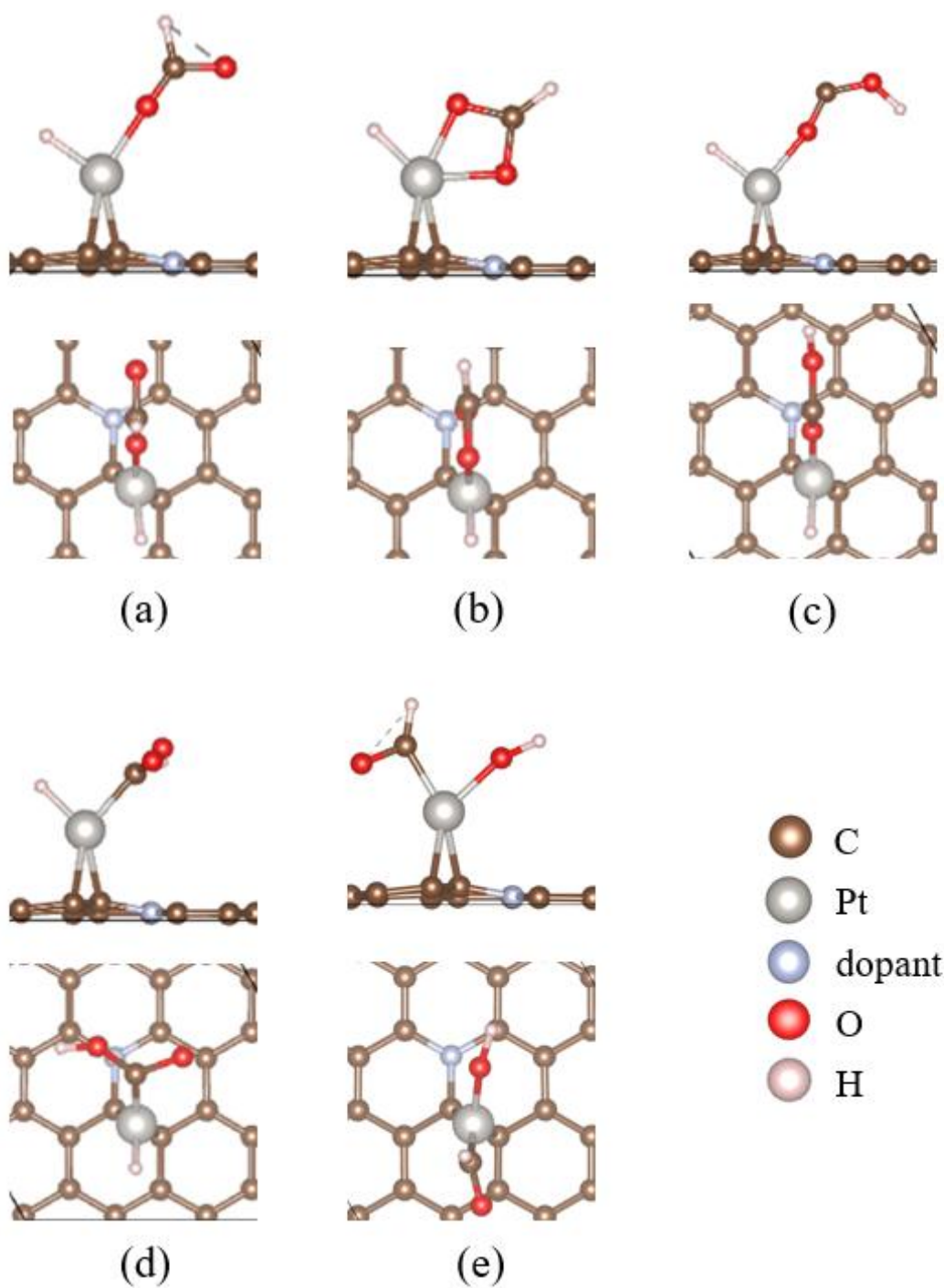

**Fig. S1.** Initial structures of intermediate adsorption states (IM) for structural relaxation. In the formate pathway, (a) monodentate adsorption ($IM_{fm}$), (b) bidentate adsorption ($IM_{fb}$); in the calboxyl pathway, (c) adsorption through the O atom ($IM_{cO}$), (d) adsorption through the C atom ($IM_{cC}$) of the carboxyl group; and in the formyl pathway, (e) one configuration ($IM_f$).

**Table S1.** Relative energy at each reaction step for the decomposition reaction of formic acid on single Pt atoms supported on pristine, N-doped, P-doped, and O-doped graphene.

|     |     | $E_r$ (eV) | | | |
| --- | --- | --- | --- | --- | --- |
|     |     | pristine | N-doped | P-doped | O-doped |
| IS  |     | -1.64 | -1.38 | -1.20 | -1.04 |
|     | $IM_{fm}$ | -1.13 | - | -1.16 | -0.92 |
|     | $IM_{fb}$ | -1.86 | -1.65 | -1.62 | -1.23 |
| IM  | $IM_{cO}$ | - | - | - | 0.95 |
|     | $IM_{cC}$ | -2.01 | -1.76 | -0.89 | -0.59 |
|     | $IM_f$ | -1.41 | -1.10 | -0.95 | -0.53 |
|     | $FS_{H_2+CO_2}$ | -1.97 | -1.56 | -0.82 | -0.55 |
| FS  | $FS_{H_2O+CO}^{calboxyl}$ | -2.64 | -2.24 | -1.57 | -1.57 |
|     | $FS_{H_2O+CO}^{folmyl}$ | -2.64 | -2.30 | -1.57 | -1.37 |